\documentclass[a4paper,11pt]{article}
\pdfoutput=1
\usepackage{jinstpub}
\usepackage{amsmath, amsthm, amssymb}
\usepackage{textcomp}
\usepackage{graphicx}
\usepackage{subcaption}
\usepackage{sidecap}
\usepackage{fontenc}
\usepackage{pbox}
\usepackage{lineno}
\usepackage{url}
\usepackage[]{units}
\usepackage{xspace}
\usepackage{gensymb}
\usepackage[htt]{hyphenat}

\sidecaptionvpos{figure}{c}

\title{Detecting Cherenkov Light From 1-2 MeV Electrons in Linear Alkylbenzene}

\author[a,1]{J.~Gruszko\note{Corresponding author.},}
\author[b]{B.~Naranjo,}
\author[c]{B.~Daniel,}
\author[d]{A.~Elagin,}
\author[a,e]{D.~Gooding,}
\author[e]{C.~Grant,}
\author[a]{J.~Ouellet,}
\author[a]{and L.~Winslow}

\affiliation[a]{Massachusetts Institute of Technology, Department of Physics, 77 Massachusetts Avenue, Cambridge, MA 02139, USA}
\affiliation[b]{University of California, Los Angeles, Department of Physics and Astronomy, 475 Portola Plaza, Los Angeles, CA 90095, USA}
\affiliation[c]{Yale University, Department of Physics, 217 Prospect Street, New Haven, CT 06511, USA}
\affiliation[d]{Enrico Fermi Institute, University of Chicago, 5640 S. Ellis Ave, Chicago, IL, 60637}
\affiliation[e]{Boston University, Department of Physics, 590 Commonwealth Avenue
Boston, MA 02215, USA}

\emailAdd{jgruszko@mit.edu}
 
\abstract{The FlatDot detector has been used to demonstrate the separation of Cherenkov and scintillation light for 1 to 2\,MeV electrons in linear alkylbenzene (LAB). With an average PMT transit time spread (TTS) of 200\,ps, the early light in each event is clearly dominated by the Cherenkov signal, which on average comprises $86^{+2}_{-3}\%$ of the light collected in the first 4.1\,ns of each event. The spatial distributions of the Cherenkov and scintillation light are found to match those predicted in Monte Carlo simulations. This is a key step towards demonstrating direction reconstruction of $\beta$ decays, a technique that could reduce $^8$B solar neutrino backgrounds for neutrinoless double-beta decay experiments in liquid scintillator.}


\newcommand{\nonubb}{$0\nu\beta\beta$}

\begin{document}
\maketitle
\flushbottom

\section{Introduction}
Searches for neutrinoless double-beta decay ($0\nu\beta\beta$) test the Majorana nature of the neutrino. If observed, this rare decay would indicate that neutrinos are Majorana particles and that lepton number is not a conserved symmetry, which would have significant implications for the origin of the neutrino mass and for theories concerning the matter/anti-matter asymmetry of the universe. For reviews, see Refs.~\cite{HenningReview, DellOroReview, SchechterValle}. The final state of this decay features two electrons emitted with 1-2\,MeV of energy each, depending on which of the dozen \nonubb\ candidate isotopes is chosen.

Given the extremely long half-life lower limits presently on this decay ($10^{25}$ to $10^{26}$ yrs), experiments searching for \nonubb\ must have large source masses and low backgrounds, with the next generation of experiments aiming to instrument 1\,ton of double-beta decay source isotope \cite{Agostini:2017jim}. Liquid scintillators are a promising candidate technology; indeed, the highest sensitivity in the current generation of experiments has been shown by KamLAND-Zen, a detector that uses $^{136}$Xe-loaded liquid scintillator \cite{KLZ2016}.

A significant challenge facing the next generation of such detectors is the impact of so-called ``irreducible'' backgrounds, i.e. those that scale with the volume of the detector, rather than with its surface area, and cannot be reduced via shielding. The most concerning of these is the background from scattering of $^8$B solar neutrinos, which produces an energetic single electron. If, however, the topology of electron tracks in the liquid scintillator can be reconstructed, these events can be distinguished from the two emitted electrons of double-beta decay and the contamination from this background source can be reduced. 

Isolation of the Cherenkov signal from the scintillation light produced by charged particles in the scintillating medium allows this directional reconstruction in addition to energy reconstruction. While there are techniques that rely purely on scintillation light to reconstruct high energy muon tracks of a few GeV in liquid scintillator \cite{Learned2009, Wonsak}, the directional information carried by Cherenkov light is particularly important for track reconstruction of low energy single-MeV electrons, where the track length is very short. Simulations have shown that reconstruction of low energy electron tracks in large liquid scintillator detectors is achievable, and may significantly improve the sensitivity of $0\nu\beta\beta$ searches \cite{direction2014,harmonics2017}. 

Previous works have shown successful Cherenkov/scintillation light separation for cosmic-ray muons in linear alkylbenzene (LAB) and an LAB scintillator cocktail containing 2,5-Diphenyloxazole (PPO) \cite{Li2016, chessEPJC, chessPRC}. Though these experiments operate in the minimum ionizing particle regime, the muons themselves have energies over 1\,GeV, and do not experience significant scattering. This work represents the first time this technique has been demonstrated for electrons with 1 to 2 MeV energies. Relative to cosmic-ray muons, these particles present the added Cherenkov signal reconstruction challenges of increased scattering, varying energy, and lower Cherenkov light yields. Though we demonstrate Cherenkov light detection only on average and in pure LAB, this is a step towards demonstrating the utility of timing-based Cherenkov/scintillation separation for background rejection in double-beta decay experiments. 

NuDot is a 1-ton liquid scintillator experiment that plans to demonstrate this background rejection technique in a full-sized detector. Construction is planned to begin in early 2019, and  will use a combination of the fast-timing detectors utilized in this work and 8-inch-diameter photomulitplier tubes (PMTs). This measurement also serves to validate the calibration techniques, detectors, and data acquisition system that are planned for use in NuDot.

In this work, we demonstrate separation of Cherenkov and scintillation signals in 1 to 2\,MeV electrons. In section~\ref{sec:det} we describe the detector and source geometry used for these measurements, and describe the calibration procedures. We show the expected detector response based on simulations in section~\ref{sec:sims}, describe the data taken and event selection procedure in section~\ref{sec:data}, and show the separation results in LAB in section~\ref{sec:analysis}. The conclusions and prospects for a larger-scale experiment are discussed in section~\ref{sec:conclusion}.

\section{The FlatDot Experiment}\label{sec:det}
\subsection{The FlatDot Detector}
The FlatDot detector is a test-stand setup of small, fast-timing PMTs designed to demonstrate the separation of Cherenkov and scintillation signals in a variety of liquid scintillator cocktails. In this study, it is used with a collimated electron source.

\begin{figure}
\begin{subfigure}[b]{0.45\textwidth}
\begin{center}
\includegraphics[width=\textwidth]{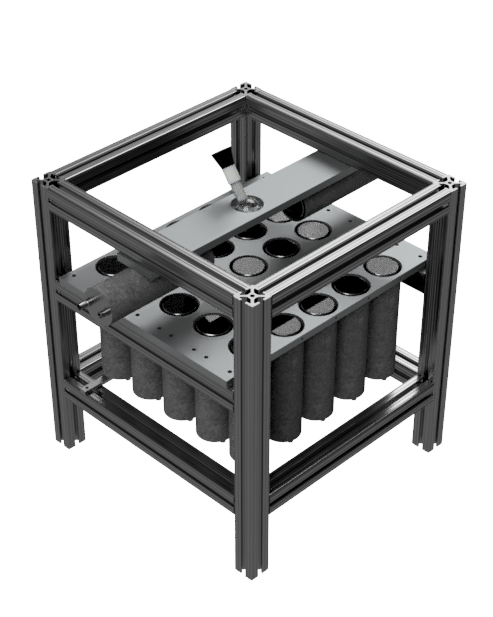}
\caption{The FlatDot detector features a 5 by 5 array of imaging (upwards-pointing) PMTs and two (horizontally-pointing) trigger PMTs, with a collimated $\beta$ decay source, at center. The source position and incidence angle can be varied as described in the text. A muon veto panel (not shown) sits above the cuvette. \label{fig:FlatDot}}
\end{center}
\end{subfigure}
~
~
\begin{subfigure}[b]{0.45\textwidth}
\begin{center}
\includegraphics[width= \textwidth]{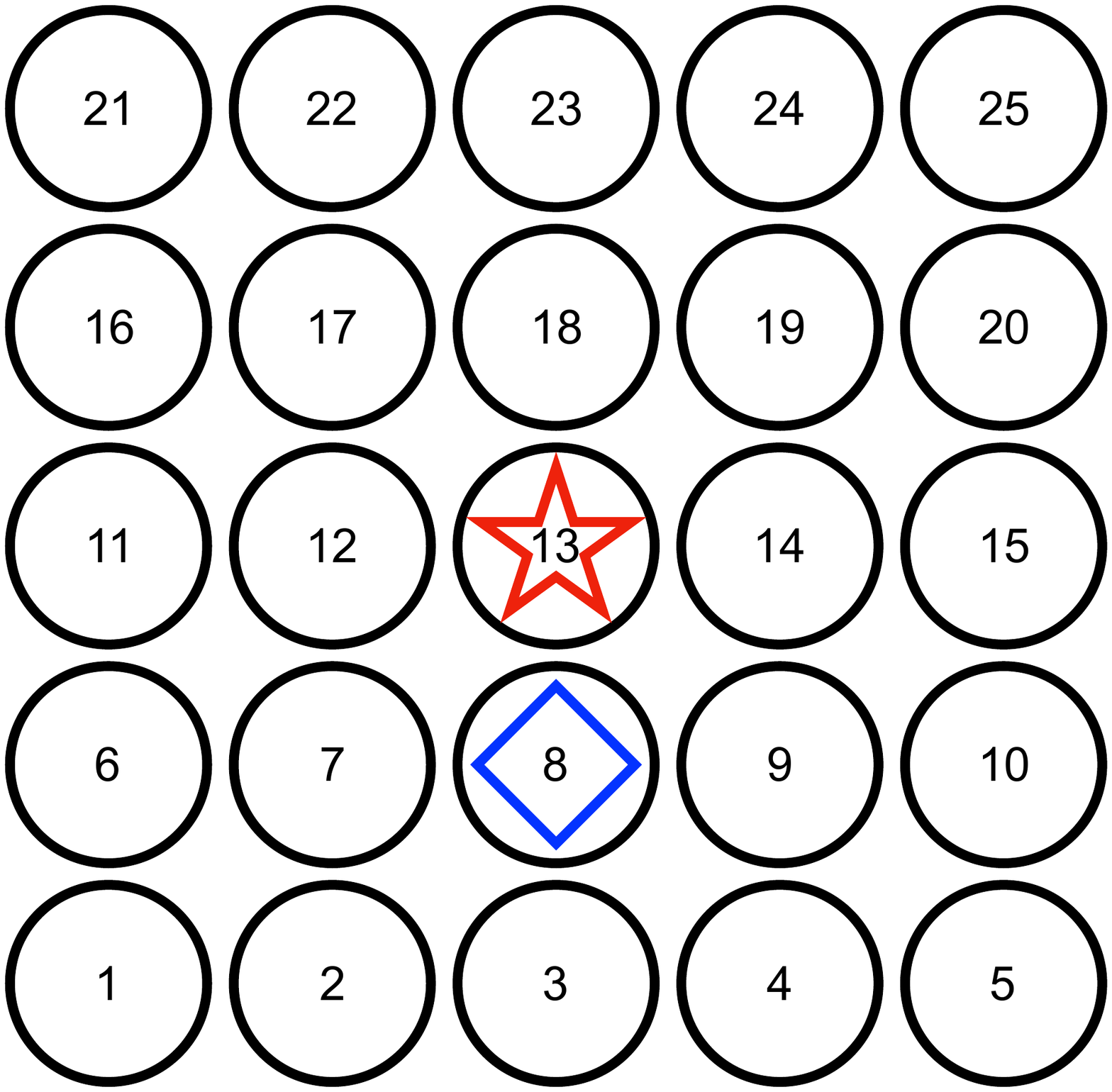}
\caption{The layout of the FlatDot array PMTs, as seen from above. In all runs, the source direction is tilted by 35$\degree$ from the normal direction to the array. In the water and "centered source" runs, the cuvette is placed above PMT 13, at the position indicated by the red star, and tilted so that the collimated source points towards PMT 8. In the "shifted source" runs. the cuvette is placed above PMT 8, at the position indicated by the blue diamond, and tilted to point towards PMT 3.}\label{fig:layout}
\end{center}
\end{subfigure}

\caption{The FlatDot experiment layout.}
\end{figure}

FlatDot employs 25 52-mm-diameter Hammamtsu Model R13089 PMTs, which are arranged in a 5 by 5 planar array spanning 1265 cm$^2$. In this paper, these are termed the ``array PMTs.'' A spherical quartz cuvette holding the liquid scintillator cocktail and collimated $^{90}$Sr source is suspended above the plane of the array, with the liquid level 16\,cm above the array. The position of the cuvette and collimated source assembly can freely adjusted in the direction normal to the plane of the array, with a minimum liquid level height of 5\,cm and a maximum height of 30\,cm. The cuvette and source assembly can also be moved to within 5\,cm of the array's edges in the lateral directions. The incidence angle of the electrons can be varied freely between 0 and 40\degree, where the angle is taken with respect to the normal axis, with any choice of azimuthal angle. 

Two additional PMTs are used to trigger the readout of the imaging array. These ``trigger PMTs'' are placed on opposite sides of the cuvette, as seen in figure~\ref{fig:FlatDot}. A coincidence of both PMTs is required to trigger the data acquisition. 

Additionally, a 9\,cm by 40\,cm plastic scintillator muon veto panel is instrumented with two veto PMTs, and placed 12\,cm above the liquid level in the cuvette. The panel is aligned with the cuvette position, and spans a row of the array PMTs. Events in which the two combined veto PMTs see a signal are vetoed at the analysis stage. 

\begin{SCfigure}
\begin{centering}
\includegraphics[width=.5\textwidth]{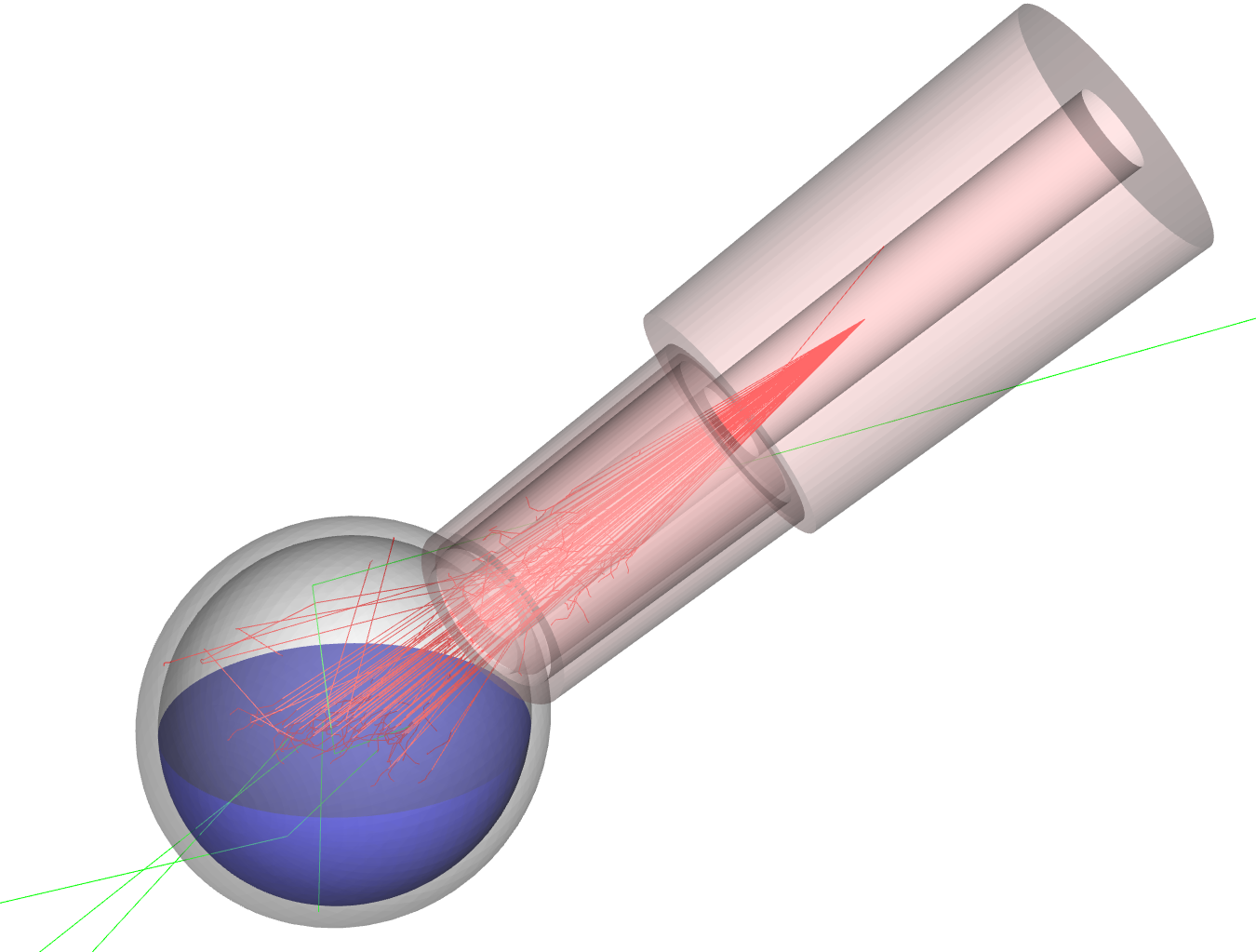}
\caption[]{A simulation of electrons in the collimator, in red. The source is inserted into the collimator bore and can be paired with a variety of target material liquids (shown in blue), held in the quartz cuvette. Cherenkov emission is not depicted in this image, but x-rays are visible, in green. In this simulation, the electrons are produced with a half-angle of 8\degree; the collimator reduces the half-angle to 5\degree. \label{fig:source_MC}}
\end{centering}
\end{SCfigure}

For these measurements, a collimated 0.01\,$\mu$Ci $^{90}$Sr needle source produced by United Nuclear was used. $^{90}$Sr decays to $^{90}$Y via $\beta$ decay with a Q value of 0.546\,MeV. $^{90}$Y subsequently decays to $^{90}$Zr via $\beta$ decay with a Q value of 2.280\,MeV in 99.99\% of decays. The electron Cherenkov threshold in LAB (given an index of refraction of 1.49 at 500\,nm \cite{reno2010LAB}) is 178\,keV; many of the initial $^{90}$Sr decay electrons fall below this threshold, and those that lie above it emit relatively few Cherenkov photons before falling below threshold. With an additional total energy cut employed at the analysis stage, the sample analyzed is expected to be a nearly-pure sample of $^{90}$Y $\beta$ decays. 

The source is mounted in a 3D-printed Nylon collimator with a long bore. The collimator is held in the neck of the cuvette, as shown in figure~\ref{fig:source_MC}, and the cuvette is filled to its midline with the liquid scintillator cocktail. The collimator used for these measurements has a half-angle opening of 5\degree, as determined using a Monte Carlo simulation. See section~\ref{sec:sims} for details of the simulation. The collimator design does not completely ensure that the needle source is centered in the bore, which leads to a few-degree uncertainty in the source position. Future versions of the collimator will include further constraints on the source position to eliminate this uncertainty.

The overwhelming majority of vertices associated with the Cherenkov cones for electrons with 1 to 2 MeV of energy occur within 1\,cm of the surface of the liquid scintillator. A 1.4\,MeV electron, for instance travels a total path length of 0.8\,cm, has a distance from the origin of 0.6\,cm in $0.030 \pm 0.004$\,ns and takes $0.028 \pm 0.004$\,ns to drop below Cherenkov threshold~\cite{direction2014}. The scattering follows the same pattern. Due to scattering, the electron's final direction does not match its inital direction. However, simulations have shown that the scattering angle is small while the majority of the Cherenkov light is produced, and the Cherenkov light encodes the direction of the primary electron at energies relevant for \nonubb. See Ref.~\cite{direction2014} and figure~7 therein.

\subsection{Data Acquisition and Triggering}
All of the array PMTs are operated at 1700\,V of bias, with the trigger PMTs held at 1750\,V. The high voltage to all PMTs is supplied by a Wiener EDS 30330n ISEG MPOD high voltage module, which has ripple below 20\,mV. Signals from the FlatDot array are digitized using a single CAEN V1742 digitizer, a 32+2 channel switched capacitor digitizer that is based on the DRS4 chip. It is set to sample at 5\,GHz, resulting in an acquisition window of 204.8\,ns. The trigger-induced dead time of this system is 110\,$\mu$s.

In the Cherenkov ring measurement setup, the signals of the two trigger PMTs are amplified by a factor of 10 in an MMIC fixed-gain low-noise amplifier. These amplified signals are each used to produce a 10\,ns-wide NIM logic gate; The coincidence of these gates, produced by a CAEN V976 FIFO logic unit, triggers the readout of the array using the V1742's low-latency local trigger option. The unamplified trigger and muon veto channels are also digitized. In the offline analysis, they are used for a more stringent (0.2\,ns) coincidence requirement and to veto cosmic muons, respectively. 

\subsection{Pulse Shape Analysis}\label{ssec:PSA}
Single photoelectron pulses in the array PMTs are 6\,ns wide. Pulses are identified using an algorithm similar to that described in \cite{DEAP2014}:
\begin{itemize}
\item a 3\,ns (15 sample) sliding average window is used to filter the waveform
\item regions in which the sliding window integral exceeds 5 times the RMS of the baseline times the square root of the number of samples in the averaging window are identified as pulse regions
\item the margins of the pulse region are extended until the point where the following 5 samples fall below the RMS of the baseline
\end{itemize}
For this analysis, the RMS of the digitizer baseline noise is taken to constant at  0.5\,mV, rather than being measured from each pulse. The integral in the pulse region is taken to be the total charge of the pulse.

The time associated with the pulse is found using a constant fraction discriminator (CFD) within the pulse region. The fraction used is 40\% of the pulse height, with a 2\,ns delay. This trigger time is used for the calibrations and the event alignment. The fraction was varied to test the optimization of the CFD trigger timing. The transit time spread of some of the PMTs can be reduced by changing the fraction to 60\%, but the effect is not uniform, with different PMTs having different optimal CFD fractions. In each case, the timing improvement gained by optimizing the CFD on a detector-by-detector basis is less than 10\%, with almost all improving by less than 5\%. Future analyses will use such an optimization strategy, but in this analysis, a uniform CFD level of 40\% is used. 

\subsection{Gain Calibration}
A pulsed LED source, mounted above FlatDot, is used to find the single photoelectron (PE) gain and pulse height uncertainty for each array PMT. The LED and digitization are both triggered using a signal generator, with the LED voltage set to produce occasional single-PE-level pulses. 

Using the approach described in section~\ref{ssec:PSA}, the total charge distribution of each PMT for single PE pulses is drawn and fit with a Gamma distribution described by shape parameters $\alpha$ and $\beta$. The mean of the distribution $\langle Q \rangle = \frac{\alpha}{\beta}$ is used to determine the gain of the PMT. This calibration is used to convert the pulse integrals to a number of photoelectrons in the subsequent analyses. The associated uncertainty is given by $\frac{\sigma}{\mu} = \frac{1}{\sqrt{\alpha}}$. The average gain of the PMTs used is $1.152 \times 10^6$, with an associated uncertainty of 30\%.   

\subsection{Timing Calibration}\label{sec:timing_calib}
As addressed in detail in Ref.~\cite{direction2014}, four main effects contribute to the timing of a scintillator detector system: the travel time of the particle, the time constants of the scintillation process, chromatic dispersion, and the timing of the photodetector. Of these, only the first and last are addressed by the FlatDot timing calibration. The time constant of the scintillator is one of the elements under study in this paper; no calibration is used to account for it. The chromatic dispersion has a minimal effect given the small distance the photons travel in liquid scintillator. The electron travel time is small, less than 30\,ps, so the timing of the photodetector dominates in the FlatDot geometry. 

\begin{figure}
\begin{center}
\includegraphics[width= .7\linewidth]{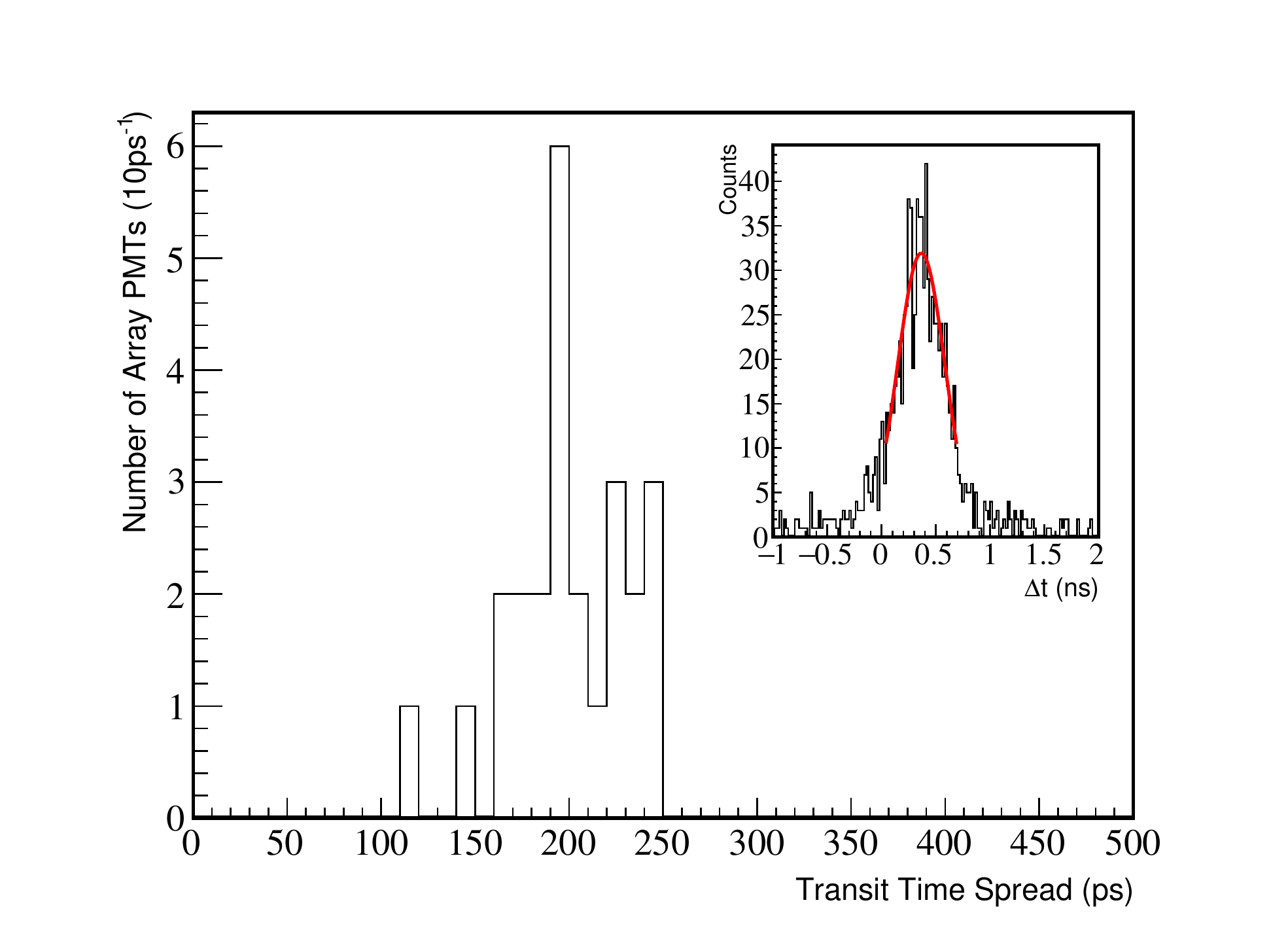}
\caption{The transit time spread (TTS) of each of the 25 array PMTs. \textit{Inset:} The time difference in triggers between PMTs 12 and 13 (the center PMT of the array). The coincident resolving time is given by the width $\sigma$ of the Gaussian fit, and the timing calibration offset is given by its centroid. The fit range is truncated to avoid the tails caused by triggering on higher-noise waveforms. \label{fig:TTS}}
\end{center}
\end{figure}

The timing calibration is determined from the Cherenkov signal produced by pairing the collimated $^{90}$Sr source with a water-filled cuvette. Given the small path length of the electron, all of the photons produced can be assumed to originate simultaneously and at a single point near the center of the cuvette, with the small effect of deviations from that assumption folded into the transit time spread. The differences in arrival times of signals in each PMT relative to a reference array PMT are fit with a Gaussian curve, as shown in the inset of figure~\ref{fig:TTS}. The timing correction needed to account for the photon travel time to each array PMT as well as any differences in the electronics is determined from the centroid of the Gaussian. Throughout the analysis, the timing of each PMT signal is corrected by this shift. Any FlatDot PMT can be chosen as the reference detector at the analysis stage; we select PMTs near the center of the array, which have the highest rate, to maximize the true coincidence rate with the other PMTs. 

The time spread of the signal, which is dominated by the transit time spread (TTS) of the PMT, is determined from the width of the Gaussian. The Gaussian width can be expressed as the coincident resolving time, $\sqrt{(\tau_1^2+\tau_2^2)}$, where $\tau_{1, 2}$ are the TTS of the each of the two PMTs \cite{HamamatsuHandbook}. Performing this procedure with two different array PMTs as the reference detector, we can extract the true TTS of each PMT. The TTS of the PMTs improves as the number of PEs per pulse increases; given the small number of expected Chernekov PEs, we include only events with 0.5 to 2.5 measured PEs per PMT in the TTS determination. 

The average TTS of the 25 array PMTs is 200\,ps, with all PMTs having TTS between 117 and 250\,ps (see figure~\ref{fig:TTS}). This value sets the best-possible Cherenkov/scintillation separation possible with these detectors, assuming there is no chromatic dispersion in the scintillator.

\section{Simulations}\label{sec:sims}
We used the RAT-PAC simulation and analysis framework \cite{RAT}, based on Geant4.10 \cite{geant4one}, 
to model the FlatDot geometry, with the physics list \texttt{QGSP\_BERT}. We simulated the emission of electrons from the assumed location of the needle source, which are then collimated by the nylon tube. The collimated electrons then impinge on the liquid scintillator inside the quartz cuvette.  

\begin{figure}[t]
\begin{subfigure}[b]{0.45\textwidth}
\begin{center}
\includegraphics[width=\textwidth]{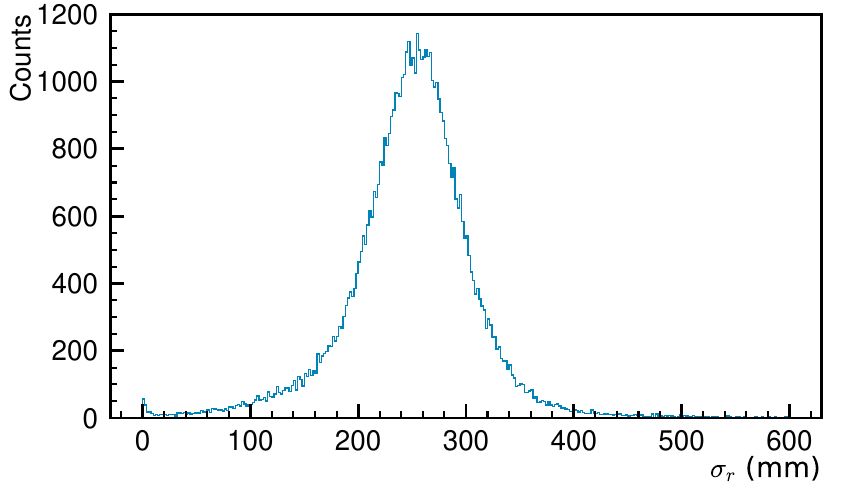}
\caption[]{The Cherenkov ring radius of the individual simulated events. The expected value of the radius is $250 \pm 61$\,mm. When the average of the events is taken, the individual rings are blurred into an asymmetric spot, as shown to to the right. \label{fig:ring_r}}
\end{center}
\end{subfigure}
~
~
\begin{subfigure}[b]{0.45\textwidth}
\begin{center}
\includegraphics[trim={0 .78cm 0 0},clip, width=\textwidth]{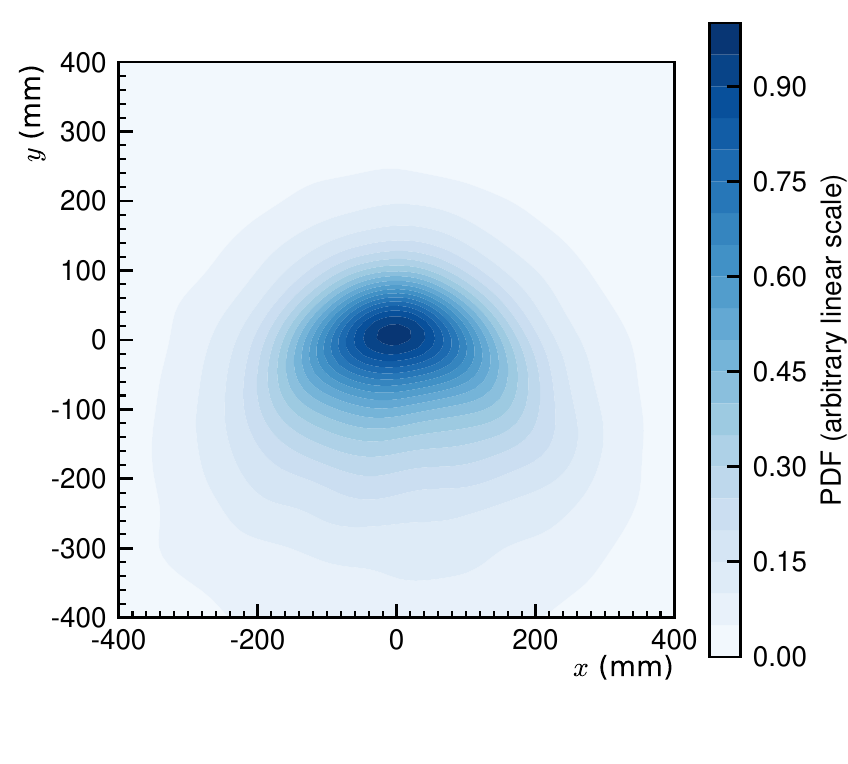}
\caption[]{The average Cherenkov cone hit pattern for the centered and tilted source, without accounting for the separate array PMTs and the space between them. 1000 events in which at least one Cherenkov photoelectron is produced are included in the average. \label{fig:Cherenkov_unbinned}}
\end{center}
\end{subfigure}
\caption{Monte Carlo simulations in RAT-PAC are used to study the propagation of electrons in the collimator and source, the production of Cherenkov and scintillation light in the cuvette, the propagation of photons to the imaging plane, and their detection. The electrons are required to have initial energy above 1\,MeV. The quantum efficiency of the imaging plane is set to match that of the R13089 PMTs.}
\end{figure}

RAT-PAC includes the full spectral shape of $^{90}$Sr and $^{90}$Y $\beta$ decay emission; the sum of these is used. A cut to include only electrons above 1\,MeV is taken, mimicking our data analysis strategy, which eliminates events from the decay of $^{90}$Sr. Cherenkov light is handled by \texttt{G4Ceren\-kov}. The Cherenkov ring formation depends on the particle direction and is therefore sensitive to perturbations from the electron's initial trajectory. \texttt{G4EmStandardPhysics}, which chooses between several scattering models for a given particle type and energy, is used to model multiple Coulomb scattering. 

In these runs, the source was tilted to give a clearly discernible distinction between the spatial distribution of Cherenkov and scintillation light. Other source orientations were also tested, but were not studied in detail. Because the source is not positioned normal to the array, the radius of the Cherenkov ring produced by each electron is best expressed as $\sigma_r = \sqrt{\sigma_x^2 + \sigma_y^2}$, where 
$$\sigma_{x} = \sqrt{\frac{1}{N-1} \sum_{i}(x_i-\bar{x})^2}$$ 
and the equivalent expression is used for $\sigma_y$. The $x_i$ and $y_i$ are the positions of each single-photoelectron hit (with each PE of a multi-PE hit counted separately), $\bar{x}$ and $\bar{y}$ are the mean positions, and $N$ is the total number of hits. 

Including the effects of scattering and the realistic wavelength-dependent quantum efficiency of the R13089 PMTs (35\% at 400\,nm), the expected radius of the ring is $250 \pm 61$\,mm, with the distribution shown in figure~\ref{fig:ring_r}. The average distribution of Cherenkov light for a centered source that is tilted by 35$\degree$ is shown in figure~\ref{fig:Cherenkov_unbinned}.

The generation of scintillation light is handled by \texttt{GLG4Scint}, with the assumed LAB emission spectrum and light yield taken to be that in Ref.~\cite{Li2016}. As in the Cherenkov light simulations, the effects of electron and photon scattering and the realistic quantum efficiency of the detectors is taken into account. 

To calculate the average charge per PMT for both the Cherenkov and scintillation signals, as shown in figure~\ref{fig:flatdot}, the finely-binned results of the simulations are integrated across the face positions of each of the 25 PMTs. 

\section{Data Taken}\label{sec:data}
\subsection{Water} 
Measurements of Cherenkov light taken with the collimated $^{90}$Sr source and a water-filled cuvette are used to determine our timing calibration, as described in section~\ref{sec:timing_calib}. In addition, these data provide us with a purely-Cherenkov sample that can be studied to understand the PMT response and included as a component in the fit of the average waveforms. 

Runs were taken with the collimated source positioned above the center of the array and tilted by 35$\degree$ from normal to the array plane, pointed towards the center PMT of the second row from the front of the array. Events used in the water Cherenkov average waveform are chosen to have a total charge between 2 and 42 PE, integrated over the whole array. They are also required to have a triggered pulse in the imaging array within the first 4.1\,ns of the event, as defined relative to the first detected pulse, whether in the array or trigger PMTs. The live time and number of events selected by each stage of the analysis are given in Table~\ref{tab:data}.

\begin{table}[t]
\centering
\begin{tabular}{l p{15mm} p{16mm} p{15mm} p{20mm} p{20mm} p{20mm}}
\hline
Medium & Source \newline Position & Live Time \newline (days) & DAQ \newline Triggers &  Events after \newline $\mu$ and trigger selection & Events with \newline Cherenkov & Events without Cherenkov\\  \hline
Water & Centered & 3.2554 & 53098  & 5891 &  2983 & N/A \\
LAB & Centered & 2.9681 & 185583 & 10318 & 4216  & 220 \\
LAB & Shifted & 1.9452 & 123346 & 8031 & 2874 & 152 \\
\end{tabular}
\caption{The data sets used in studying the FlatDot response. The live time for each data set takes into account the 110\,$\mu$s DAQ dead time induced by each trigger. The last two columns give the number of events used in building each of the average waveforms. For each event in the average, 25 waveforms (one per array PMT) are used to build the array-average waveforms.} \label{tab:data}
\end{table}

\subsection{Linear Alkylbenzene}
Events composed of a combination of Cherenkov and scintillation light are studied using the collimated $^{90}$Sr source and cuvette filled with pure LAB. Data were taken in two configurations, both with the collimator tilted as in the water runs. 

The centered-source runs were taken with the cuvette placed above the center PMT in the array, which is in the third row of detectors. The shifted-source runs are taken with the cuvette and source shifted by one row, and placed above the center PMT in the second row of detectors. These shifted runs allow us to study a larger range of source distances from the detectors.

\begin{figure}
\begin{center}
\includegraphics[width= .7\linewidth]{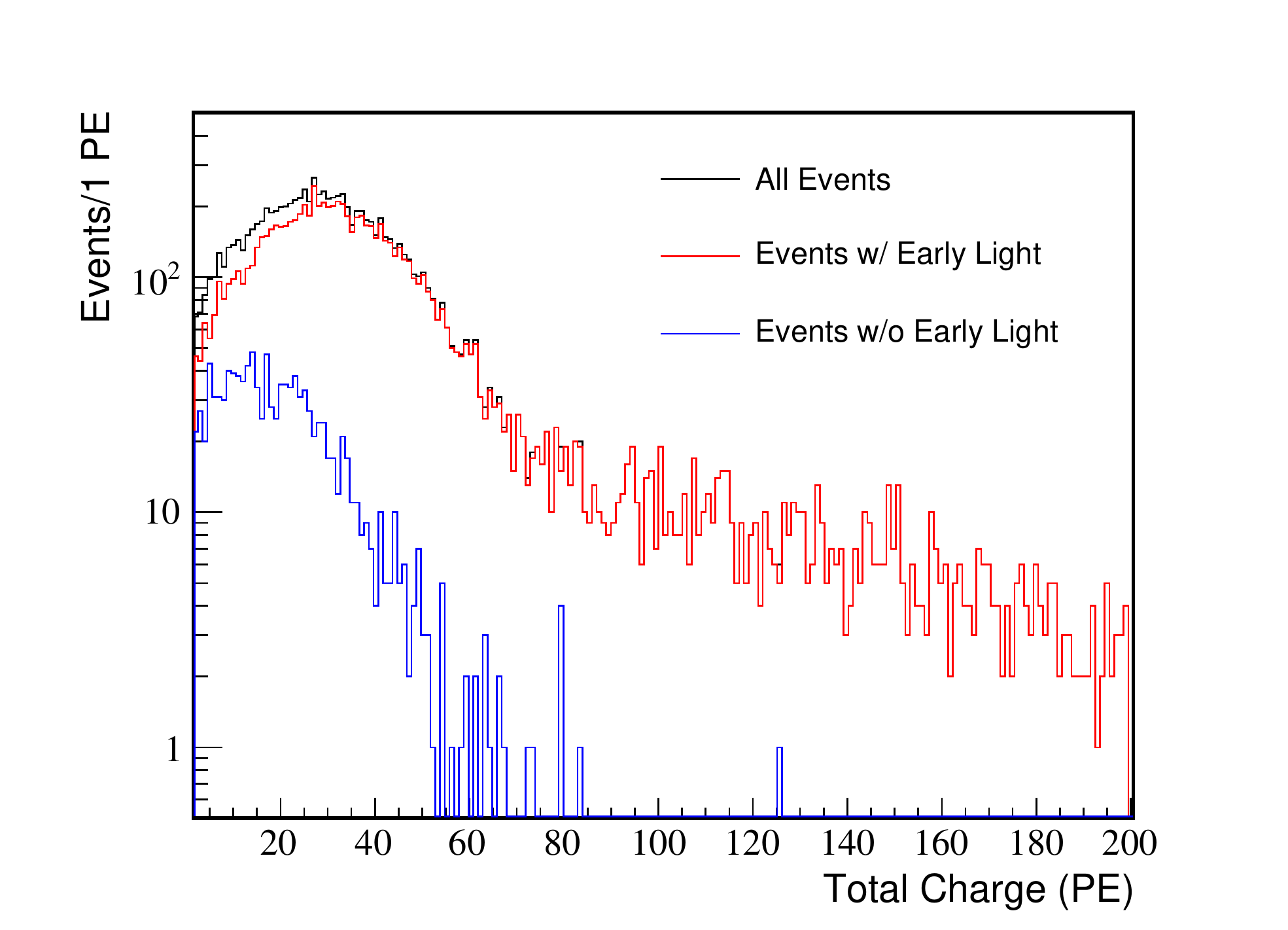}
\caption{The energy spectrum in the centered-source LAB runs. Events passing the initial trigger coincidence and muon veto selection are in black, with accepted Cherenkov-containing events in red. Events not containing early array hits are in blue. Events with 30 to 75\,PE of charge in the array are used for this analysis. Events in the high-energy tail are cosmic muons that fail to pass through the veto panel. \label{fig:spec}}
\end{center}
\end{figure}

The spectrum in the centered-source run can be seen in figure~\ref{fig:spec}. In all LAB data sets, we require accepted events to have between 30 and 75 PE, so that they lie in the upper half of the $^{90}$Y $\beta$ spectrum, above approximately 1\,MeV. The use of a high-energy threshold helps eliminate background muon events that are not tagged by the muon veto. 

We build a Cherenkov light-containing data set, where we impose the requirement that early triggers be present in the array, as discussed below. We also build a data set of Cherenkov-free events, where we require that early triggers only be present in the trigger PMTs. The details of the resulting data sets are given in Table~\ref{tab:data}. 

\section{Analysis and Results}\label{sec:analysis}
\subsection{Event Selection and Timing Alignment}
For all analyses, we retain only events in which the two trigger PMTs have CFD-triggered pulses within 0.2\,ns of one another. We also reject all events in which the combined muon veto PMT charge exceeds one PE. The baseline voltage of each trace is calculated from the first 20\,ns (100 samples) of the waveform; the DAQ system parameters as set such that the first trigger occurs 25 to 30\,ns into the waveform. We impose the requirement that a pulse is not triggered in the baseline calculation window and that the RMS of the calculated average not exceed 0.5\,mV. The total charge of each event is determined from the summed charge of the array PMTs. We impose an upper and lower charge threshold, determined separately for the Water and LAB data sets. 

The trigger time ($t_0$) of each event is taken to be the CFD crossing time of the earliest pulse in any of the PMTs. For studies of scintillation-only events, we choose events that are triggered by one of the two trigger PMTs, with no light signals occurring in the array within 4.1\,ns of $t_0$. For studies of Cherenkov-light-containing events, we require that at least one of the array PMTs have a triggered pulse in the early time window. The 4.1\,ns early light window is determined from studies of the LAB signal time evolution, as described below. 

\subsection{Array-Average Analysis}
While the CFD-triggered approach allows us to reliably implement the basic event selection,  incurs an analysis dead time due to the requirement that the signal return to baseline before a subsequent trigger can occur. To leverage the fast response of the PMTs and their zero intrinsic dead time at low light levels, we use an average-pulse method for our analysis of the time evolution of the signal. 

\begin{figure}
\begin{center}
\includegraphics[width= .85\linewidth]{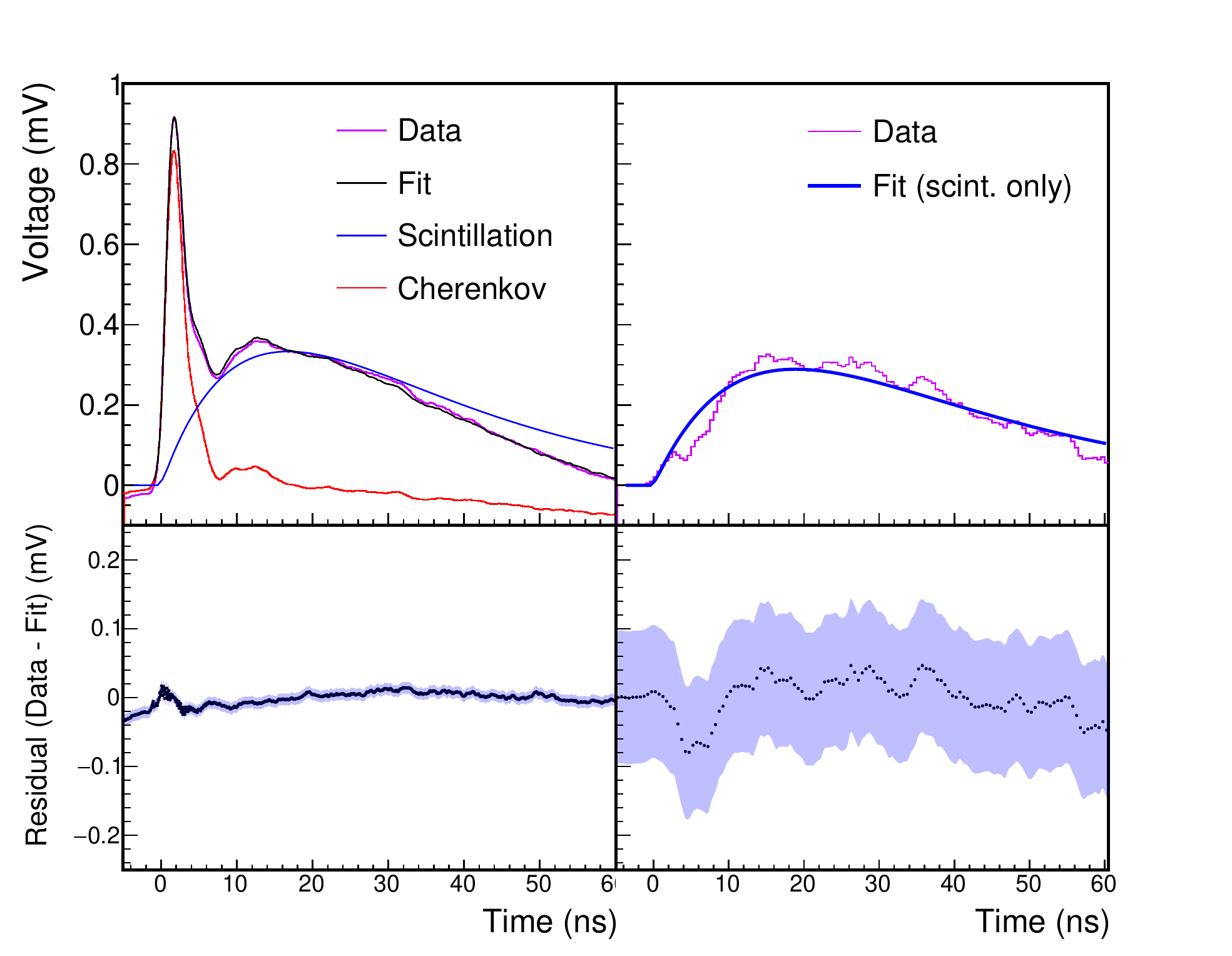}
\caption{The average waveforms and fit results from the centered-source LAB data set. \textit{Upper left: }The average waveform of 4216 events in which Cherenkov light is detected in the array. The average waveform is shown in violet, with the fit result in black. The red line shows the Cherenkov component of the fit, with the scintillation component in blue. \textit{Upper right:} The average waveform of the 220 events in which only scintillation light is detected. The fit result, composed of only the scintillation component, is shown in blue. \textit{Bottom:} The residuals of the fit for the Cherenkov-containing sample \textit{(left)} and the scintillation-only sample \textit{(right)}, with the shaded band indicating the uncertainty in the average waveform. The results and goodness-of-fit of all the fits are given in Table~\ref{tab:fitResults}. \label{fig:LAB_fit}}
\end{center}
\end{figure}

The average baseline voltage, calculated as described above, is subtracted from each digitized signal. Using the results of the timing calibration, the waveforms of each PMT are aligned with respect to one another for a single event. Linear interpolation between the sampled points is used to resample the traces as necessary and the events are all aligned with respect to $t_0$. All events passing the analysis cuts are then averaged together, and the average waveform is resampled into 50\,ps bins (20\,GHz effective sampling frequency). In the data sets that do not contain early light, which have low statistics, the average waveform requires additional smoothing-- we sample these waveforms into 500\,ps bins (2\,GHz effective sampling frequency). These average waveforms can be seen in figure~\ref{fig:LAB_fit}.

We determine the uncertainty in the average pulse via Monte Carlo simulation. The average pulse in each PMT is converted to an integrated photoelectron charge per event using the gain of the PMT, and these values are summed to give the average array charge per event. This number, multiplied by the number of events included in each average pulse, is the number of samples drawn randomly from the probability distribution created by the average pulse. The residuals of the resulting distribution are fit with a Gaussian curve, and its width is taken to be the uncertainty. This procedure is repeated 1000 times, resulting in an uncertainty of $\pm 8.5 \mu$V per sample in the average waveforms that include Cherenkov hits. Due to the lower statistics of the samples that do no contain early array hits, the associated uncertainty for these average waveforms is larger, $\pm 97 \mu$V for the centered-source runs and $\pm 94 \mu$V for the shifted-source runs. 

The resulting average waveform is fit using the sum of a scaled Cherenkov average waveform, taken from data, and a theoretical scintillation curve:
$$
V(t) = A \frac{\tau_{r}+\tau_{d}}{\tau_d^2}(1-\exp{(-t/\tau_r)})\exp{(-t/\tau_d)} + B(V_C(t)).
$$
The amplitudes A and B are allowed to float in the fit, as are $\tau_r$ and $\tau_d$, the rise and decay time constants, respectively, of the scintillator. $V_C(t)$ is the average waveform of the water Cherenkov data, normalized to a peak height of 1\,mV. 

Due to this hybrid approach, the fitting function accounts for the overshoot and afterpulse effects that are induced by the Cherenkov signal, but not those caused by the additional scintillation light. The overshoot in particular has a significant effect after $t=70$\,ns -- the latter half of each average pulse falls below zero before returning to baseline. Since this portion of the waveform contains little useful information, we truncate the fit region to $-1<t<60$\,ns. The included Cherenkov overshoot accounts for the fact that the fitted scintillation component is of higher amplitude than the Cherenkov-containing average waveform in the latter half of the fit window; the negative amplitude of the Cherenkov curve overshoot is being added to the total fit waveform. 

The theoretical scintillation curve does not account for the presence of additional decay time constants, which are expected in liquid scintillator. Their contribution is expected to be small. Due to the low scintillation light levels studied here, we do not have the capability to determine these additional parameters and they are not included in the fit. Since the model used for the scintillation curve is normalized such that the degeneracy between $\tau_r$ and the amplitude $A$ is removed, the change in the shape of the curve caused by increasing $\tau_r$ is minimal when $\tau_r$ is comparable to or longer than $\tau_d$. Given the low scintillation light levels and slow response of LAB when no additional fluor is added to it, the rise time of the scintillator is not well-constrained by the fit. 

The average waveforms of the scintillation-only events are fit using the three free parameters of the theoretical scintillation curve, with $B$ fixed at 0. If $B$ is allowed to float in this fit, its resulting value is consistent with 0. This average waveform has significant uncertainty due to the small number of such events in the selected energy region, as seen in the bottom panel of figure~\ref{fig:LAB_fit}. 

An alternative fitting approach using a scintillation average waveform taken from $\gamma$ source runs was also considered. To insure that there is no Cherenkov light contamination in the template waveform, the $\gamma$ source used must have its Compton shoulder lying below the Cherenkov threshold in LAB. $^{137}$Cs, with a $\gamma$ emission peak at 662\,keV, satisfies this requirement. Using such a source, however, the events from the $\gamma$ interactions lie far below the total array threshold of 30 PE used to analyze the $^{90}$Sr data. Given the effect of overshoot on the observed decay time of the scintillation curve, the average waveform made from these low-energy events cannot be simply rescaled to match the scintillation component expected in the $^{90}$Sr events. Therefore additional free parameters must still be used to correct the template, and this approach cannot be used to reduce the number of degrees of freedom in the fit. Given that the free parameters of the theory-driven scintillation curve are more easily interpreted than the required adjustment parameters used in the data-based scintillation fit, the hybrid approach was judged to be preferable. 

\subsection{Array-Average Results}
The result of the array-average fit in the centered-source LAB data set is depicted in the top panel of figure~\ref{fig:LAB_fit}. The same procedure is used to fit the average waveform for the shifted-source LAB data set, with the results for all fits shown in Table~\ref{tab:fitResults}. In LAB, the Cherenkov and scintillation signals are clearly separable in time, with the Cherenkov signal dominating the average pulse for the first 4.1\,ns. In this window, $86^{+2}_{-3}$\% of the charge detected is found to be due to Cherenkov light. The early time window includes 78\% of the total emitted Cherenkov light while avoiding the effects of afterpulsing in the PMTs.

In spite of the increased statistical uncertainty in the scintillation-only dataset, the decay time and amplitude of the scintillation light are similar to that found in the Cherenkov-containing sample. The decay time found is unchanged if the fit is performed without the constraint of $B = 0$. The small discrepancy between the resulting decay times is likely due to the additional time constant introduced by the Cherenkov overshoot, which only contributes to the Cherenkov-containing average waveform fits.  

\begin{table}[t]
\centering
\begin{tabular}{l r r r r r}
\hline
Data Set & A (mV) &  $\tau_r$ (ns) &  $\tau_d$ (ns) & B & $\chi^2/$dof \\  \hline
Centered source, w/ Cherenkov & 15 $\pm$ 1 & 100 $\pm$ 70  & 18  $\pm$ 2 & 0.8 $\pm$ 0.1  & 1040.05/1216 \\
Centered source, no Cherenkov & 15 $\pm$ 1 & 100 $\pm$ 48  & 21  $\pm$ 2 & N/A & 112.986/119 \\
Shifted source, w/ Cherenkov & 15 $\pm$ 1 & 100 $\pm$ 80  & 18  $\pm$ 2 & 0.9 $\pm$ 0.1 & 1376.35/1216 \\
Shifted source, no Cherenkov & 14 $\pm$ 1 & 100 $\pm$ 48  & 20  $\pm$ 2 & N/A  & 115.127/119 \\
\end{tabular}
\caption{Fit results for all LAB data sets. The rise time of the scintillator can not be determined from these measurements, as discussed above.} \label{tab:fitResults}
\end{table}

\subsection{Spatial Distribution Analysis}
To study the spatial distribution of the Cherenkov and scintillation signals, waveforms from each array PMT are averaged separately, instead of being combined. These waveforms are then fit separately, with $A$ and $B$ allowed to float. $\tau_r$ and $\tau_d$ are fixed to the array-average value, and the fit region is constrained to (-1, 35)\,ns. 

\begin{SCfigure}
\begin{centering}
\includegraphics[width= .65\linewidth]{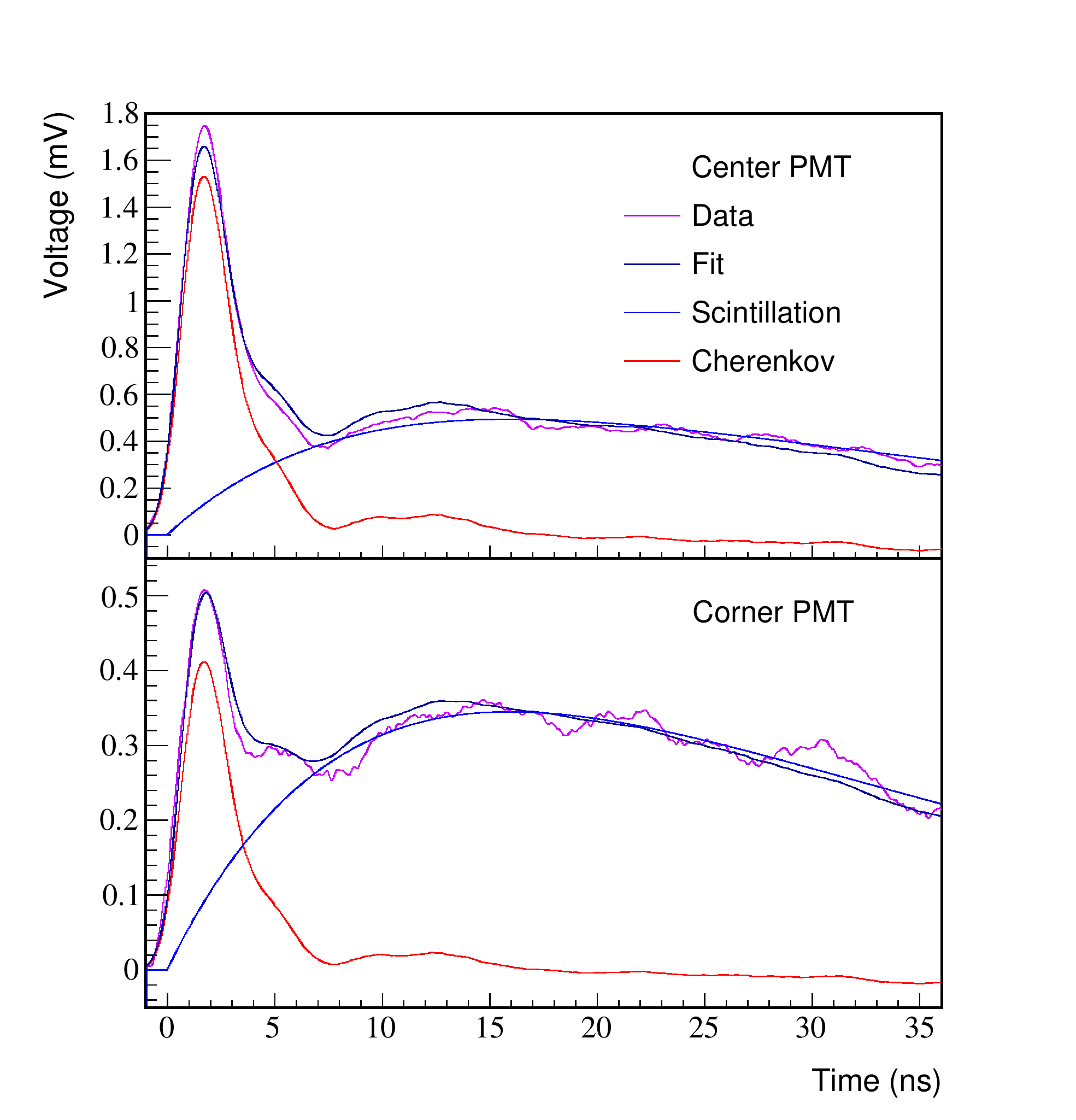}
\caption{The average waveforms and fit results from the center PMT (PMT 13, \textit{top}) and one of the corner PMTs (PMT 21, \textit{bottom}) in the centered-source LAB data set, selecting only events with visible Cherenkov hits. The average waveform appears in violet, with the fit result depicted in black. The relative fraction of the Cherenkov and scintillation components (in red and blue, respectively) vary depending on the PMT position, as do their overall amplitudes. \label{fig:2PMTs}}
\end{centering}
\end{SCfigure}

As seen in two sample PMTs (see figure~\ref{fig:2PMTs}), the Cherenkov and scintillation amplitudes vary depending on the PMT's position in the array. Integrating the components of the fit results for each PMT, we can compare the average Cherenkov and scintillation charge collected, and compare these values to the results of the simulation. To give the Cherenkov charge, the Cherenkov component is integrated in the region from -1 to 4.1\,ns; beyond this time, the majority of the signal is due to PMT ion feedback, afterpulsing, and overshoot. The scintillation charge is found from the integral of the scintillation component in the region from -1 to 150\,ns, with the long integration window chosen to make the results as comparable as possible to the Monte Carlo simulations, which place no requirement on the arrival time of the scintillation photons.

\begin{figure}
\begin{center}
\includegraphics[width= \linewidth]{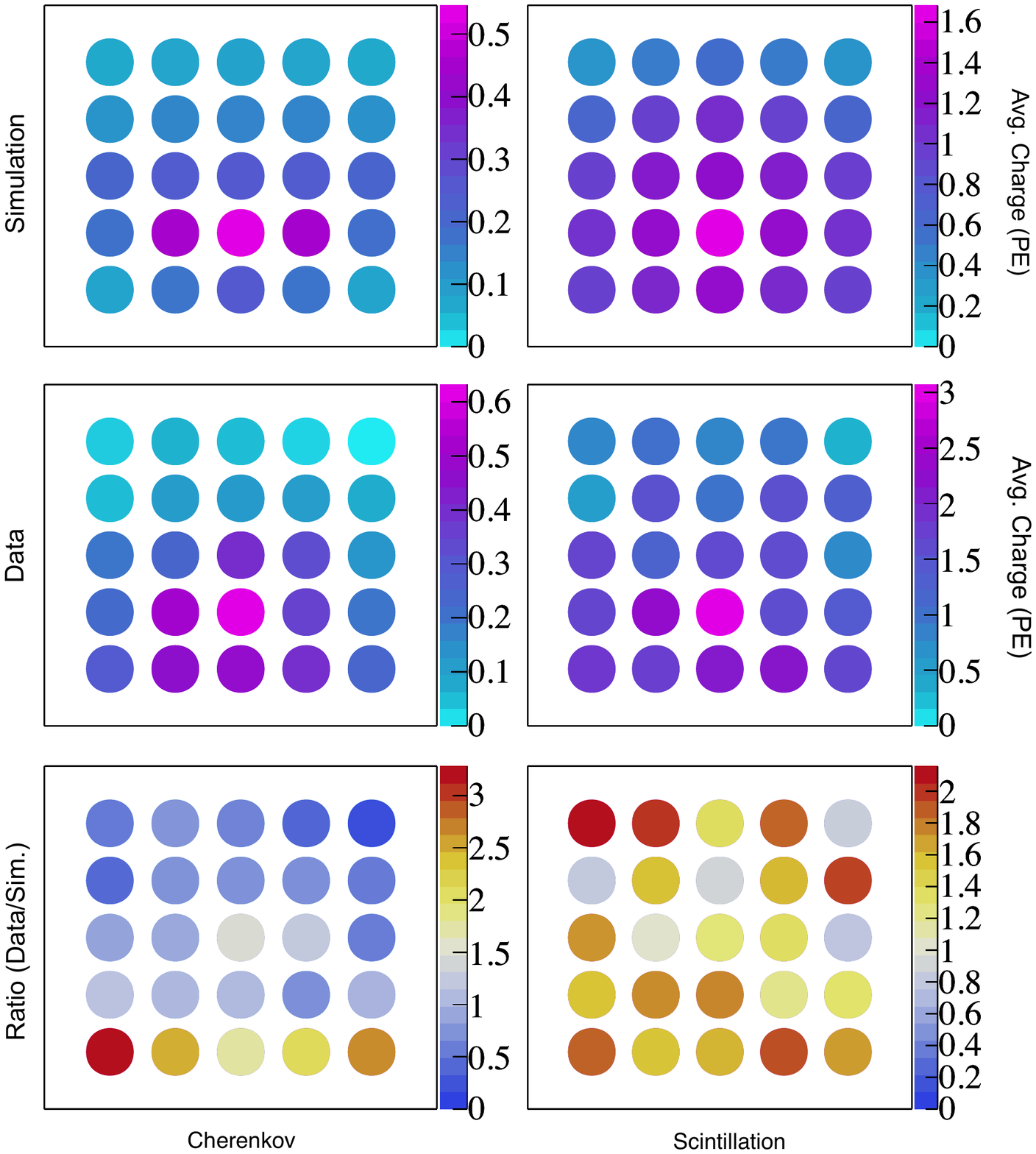}
\caption{The Cherenkov \textit{(left)} and scintillation \textit{(right)} charge spatial distributions in the FlatDot array for the shifted-source runs. In the top two rows, the color scale gives the charge, in units of photoelectrons. The simulation \textit{(top)} and data \textit{(middle)} show good agreement, save for the features discussed in the text. This can be seen in the ratio of data to simulation values \textit{(bottom)}, where the color scale indicates the value of the ratio. \label{fig:flatdot}}
\end{center}
\end{figure}

\subsection{Spatial Distribution Results}
The results from the integrated fits to the shifted-source data set are shown in figure~\ref{fig:flatdot}, and are directly comparable to the simulations of the FlatDot geometry. Some additional asymmetry appears to be present in the Cherenkov light distribution. We believe that this is due to uncertainty in the collimator angle and the fact that the collimator design does not completely constrain the source needle centering, allowing a small offset. It appears that the source in these runs may have been pointing further towards the bottom left of the array than the quoted position. Since the needle position is not completely reproducible in this collimator, this hypothesis cannot be tested until the collimator design is updated. 

The results also show a larger average scintillation signal than expected in almost all PMTs, indicating that simulated light yield may have been inaccurate. The scintillation light also appears slightly more scattered in the data than in the simulation, with PMTs at larger distances from the source showing a larger excess signal. The effects of scattering and reflections in FlatDot are under continuing study.

\begin{figure}
\begin{center}
\includegraphics[width= .8\linewidth]{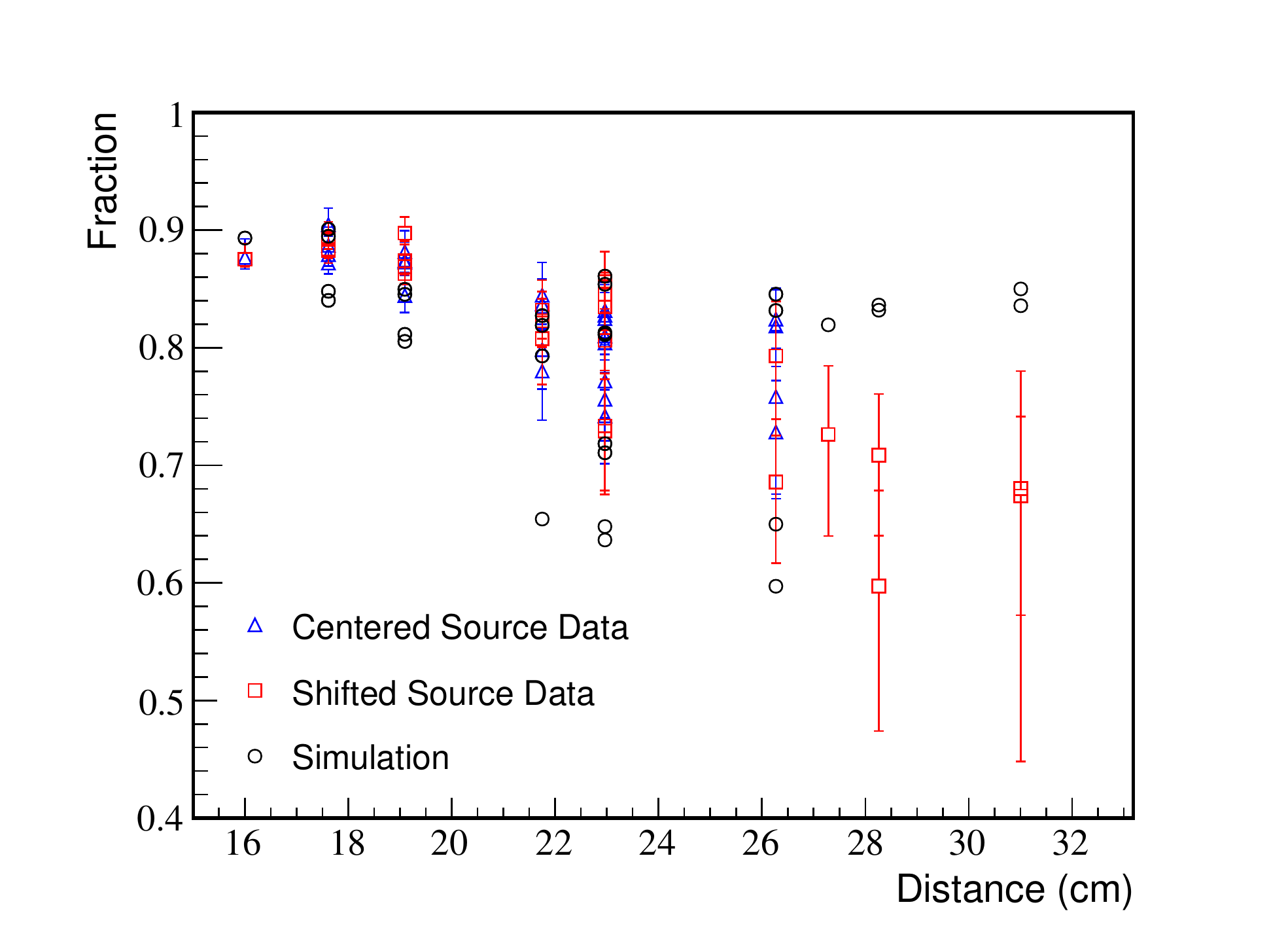}
\caption{The Cherenkov fraction of the signal in the early charge window (-1 to 4.1\,ns) falls as a function of distance from the source in both the centered-source runs \textit{(blue triangles)} and the shifted-source runs \textit{(red squares)}. In the array-average waveform, ($86^{+2}_{-3}$)\% of early light is from the Cherenkov fraction. The result of the RAT simulation \textit{(black circles)} shows disagreements consistent with the discrepancies seen in figure~\ref{fig:flatdot}. Error bars are suppressed for the simulation, since the statstical errors are negligible. The systematic error associated with the simulation will be a topic of future study. Each point represents a single PMT at a given distance from the source. \label{fig:cherFrac}}
\end{center}
\end{figure}

It is also instructive to study the Cherenkov fraction of the total early (-1 to 4.1\,ns) charge signal as a function of the PMT distance from the source, as in figure~\ref{fig:cherFrac}. This demonstrates that while both scintillation and Cherenkov light levels fall with distance, the Cherenkov light is less diffuse, as expected. In all of the detectors, over 60\% of the early light is due to the Cherenkov signal. 

There is slight disagreement between the data and simulation in the row of PMTs farthest from the source. From our interpretation of figure~\ref{fig:flatdot}, this is due to both the Cherenkov and scintillation signals; the Cherenkov light levels observed in this row were lower than expected, and the scintillation light levels were higher. 

\section{Conclusion}\label{sec:conclusion}
We find that given the timing profile of pure LAB, the Cherenkov signal dominates the average waveform for the first 4.1\,ns. In this time window, 86\% of the light is due to the Cherenkov signal, and we are able to reproduce the spatial distribution of Cherenkov light expected from our simulations. To our knowledge, this is the first such demonstration for electrons at energies relevant for $0\nu\beta\beta$ measurements, and validates several aspects of the design planned for the full-scale realization of the NuDot experiment. 

The PMTs and DAQ system planned for use in NuDot have been shown to meet the required timing performance, with an average PMT TTS of 140\,ps. We have also demonstrated the use of the collimated electron source design that will ultimately be used to calibrate and validate NuDot's Cherenkov/scintillation separation capabilities. 

Though separation remains to be shown in faster-rising and brighter scintillator cocktails, like LAB with added PPO, FlatDot's small scale and high scintillation light levels make such a measurement far more challenging than in the full NuDot design, where chromatic dispersion improves the timing separation of Cherenkov light \cite{direction2014} and the impact of reflections is expected to be minimal. Future work with both FlatDot and NuDot will move towards this demonstration. Such measurements will also allow further study of event-by-event separation, to supplement the average-event separation demonstrated here. 

Though these techniques remain to be tested, this work demonstrates a necessary first step towards the reconstruction of double-beta decay event topologies in large-scale liquid scintillator experiments. 

\subsection*{Acknowledgments}
This material is based upon work supported by the National Science Foundation under Grant Numbers  1554875 and 1806440. B. Daniel's work was funded by the MIT Summer Research Program. J. Gruszko is supported by a Pappalardo Fellowship in Physics at MIT. The work at the University of Chicago is supported by U. S. Department of Energy,Office of Science, Office of Basic Energy Sciences and Offices of High Energy Physics and Nuclear Physics under contracts DE-SC0008172 and DE-SC0015367; the National Science Foundation under grant PHY-1066014; and the Physical Sciences Division of the University of Chicago. We would like to give special thanks to Chris Haynes of MIT's International Design Center for his generosity in assisting with the 3D-printing of the collimator prototypes. We also thank Suzannah Fraker and Taritree Wongjirad for contributions to earlier iterations of this work.

\bibliographystyle{JHEP}
\bibliography{bibliography_FlatDot_2018.bib}

\end{document}